\documentstyle[12pt]{article}\textheight 230mm\textwidth 150mm
\newcommand{\be}{\begin{eqnarray}}
\newcommand{\ee}{\end{eqnarray}}
\newcommand{\eel}[1]{\label{#1}\end{eqnarray}}
\newcommand{\r}[1]{(\ref{e:#1})}
\newcommand{\vb}{{\cal h}}
\newcommand{\hb}{{\cal i}}
\newcommand{\ra}{{\rightarrow}}
\newcommand{\nn}{\nonumber}
\newcommand{\eg}{{\em e.g.\ }}

\newcommand{\al}{\alpha}

\newcommand{\la}{{\lambda}}

\newcommand{\del}{{\delta}}
\newcommand{\Om}{\Omega}
\newcommand{\pet}{{\cal P}}
\newcommand{\hpet}{\hat{{\cal P}}}
\newcommand{\heta}{\hat{\eta}}
\newcommand{\bata}{\bar{\eta}}
\newcommand{\bapet}{\bar{\pet}}

\newcommand{\bett}{{\bf 1}}
\newcommand{\halv}{\frac{1}{2}}

\begin{document}
\begin{titlepage}
\noindent
G\"{o}teborg ITP 93-19\\
August 1993\\
\vspace*{30 mm}
\begin{center}{\LARGE\bf A note on path integrals and time evolutions in BRST
quantization.}\end{center} \begin{center}\vspace*{12 mm}\begin{center}Robert
Marnelius \\
\vspace*{7 mm}
{\sl Institute of Theoretical Physics\\
Chalmers University of Technology\\
S-412 96  G\"{o}teborg, Sweden}\end{center}
\vspace*{15 mm}
\begin{abstract}
Recent formal solutions of BRST quantization on inner product
spaces within the operator method are shown to lead to an unexpected
interpretation of the
conventional path integral formulation. The relation between the Hamiltonians
in the two
formulations is nontrivial. For the operator method the correspondence requires
certain quantum rules
which make the formal solutions exact, and for the path integral the
correspondence yields a precise
 connection between boundary conditions and the choice
of gauge fixing.\end{abstract}\end{center}\end{titlepage}

\setcounter{page}{1}

The original development of the quantization procedure for gauge theories was
made within the path
integral formulation \cite{FP,BRS,FV,BV}. The operator method  came later
\cite{CF,KO} and   has
 been helpful to acquire a more precise understanding of the BRST quantization.
Recently we have
derived general formal solutions for a large class of BRST models on inner
product spaces
within the operator formulation \cite{Simple,Gauge,Propa}. As we shall see
these solutions lead to an
unexpected interpretation of the original path integral expressions. In
addition, this connection gives information of  how the formal solutions in
\cite{Simple,Gauge} are to
be quantized, rules which are in agreement  with those argued for in
\cite{Propa}. For the path
integral expressions we obtain a precise connection between the choices of
gauge fixing and the
boundary conditions to be imposed.

The conventional connection between operator quantization and path integrals is
through the
time slice formula ($\hbar=1$)
\be
&& \vb\phi',t'|\phi,t\hb=\vb\phi'|e^{-i(t'-t)H}|\phi\hb=
\int d^nq'd^nq\phi'^*(q')\phi(q)\vb q',t'|q, t\hb\nn\\
&&=\vb q',t'|q, t\hb=\int\prod_{m=1}^{N-1} d^nq_m\prod_{k=0}^{N-1}\vb
q_{k+1}, t_{k+1}|q_k, t_k\hb=\nn\\
&&=\int\prod_{m=1}^{N-1}
d^nq_m\prod_{k=0}^{N-1}\frac{d^np_k}{(2\pi)^n}\exp{(ip_k\cdot\Delta
q_k-i\Delta t H(p_k,
\bar{q}_k))}\nn\\
&&\stackrel{\lim_{N\ra\infty}}{\longrightarrow}\int_{Path}\prod_t\frac{d^nq
d^np}{(2\pi)^n}\exp{i\int^{t'}_tdt(p\cdot\dot{q}-H(p,q))}
\eel{e:1}
where $q_0=q, \;\;q_N,q_{\infty}=q', \;\;t_0=t, \;\; t_N,t_{\infty}=t',
\;\;\Delta q_k=q_{k+1}-q_k$
and $\bar{q}_k=\halv(q_{k+1}+q_k)$.  $H(p, q)$ is  the Weyl transform of the
Hamiltonian operator
$H(P, Q)$ defined by
\be
&&H(p, q)=\frac{1}{(2\pi)^n}\int d^nu d^nv \tilde{H}(u,v)e^{-i(q\cdot u+p\cdot
v)}\nn\\
&&H(P, Q)=\frac{1}{(2\pi)^n}\int d^nu d^nv \tilde{H}(u,v)e^{-i(Q\cdot u+P\cdot
v)}
\eel{e:2}
which implies
\be
&&\vb q''|H(P,Q)|q'\hb=\int\frac{d^np}{(2\pi)^n}e^{i(q''-q')\cdot
p}H(p,\bar{q})
\eel{e:3}
where $\bar{q}\equiv\halv(q''+q')$, a relation which is used in \r{1}. (The
canonical conjugate
operators $Q_i$ and $P_i$, $i=1,\ldots,n$, are hermitian and satisfy $[Q_i,
P_j]_-=i\del_{ij}$, and
$|q\hb$ are eigenstates to $Q_i$ with real eigenvalues $q_i$.)

Eqn \r{1} represents a connection between operator quantization and path
integrals for ordinary
quantum mechanics in which $|\phi, t\hb$ belongs to a Hilbert space. In a gauge
theoretic framework
one considers a larger state space which also contains indefinite metric
states. For such states
hermitian coordinate and momentum operators do not have real spectra
\cite{Pauli}.  Instead they have
imaginary ones if a Hilbert space topology is imposed (which is natural). In
such a case the
eigenstates satisfy \cite{Gross,HeT,Gen} \be &&Q|iq\hb=iq|iq\hb,\;\;\;\vb
-iq|\equiv(|iq\hb)^{\dag},\;\;\;\vb iq|iq'\hb=\del^n(q-q')\nn\\ &&\int
d^nq|iq\hb\vb iq|=\int
d^nq|-iq\hb\vb -iq|=\bett \eel{e:4}
which modifies the time slicing formula \r{1}. Instead of \r{3} we get
\be
&&\vb iq''|H(P,Q)|iq'\hb=\int\frac{d^np}{(2\pi)^n}e^{i(q''-q')\cdot p}H(-ip,
i\bar{q})
\eel{e:5}
where $H(-ip, i\bar{q})$ is a real function when written in terms of real
arguments if $H(P,Q)$ is
hermitian. The path integral formula \r{1} is then turned into
\be
&& \vb\phi',t'|\phi,t\hb=\vb\phi'|e^{-i(t'-t)H}|\phi\hb=
\int d^nq'd^nq\phi'^*(-iq')\phi(iq)\vb iq',t'|iq, t\hb
\eel{e:6}
where (see also \cite{Gross,HeT})
\be
&&\vb iq',t'|iq, t\hb=\int\prod_{m=1}^{N-1}
d^nq_m\prod_{k=0}^{N-1}\vb iq_{k+1}, t_{k+1}|iq_k, t_k\hb=\nn\\
&&=\int\prod_{m=1}^{N-1}
d^nq_m\prod_{k=0}^{N-1}\frac{d^np_k}{(2\pi)^n}\exp{(ip_k\cdot\Delta
q_k-i\Delta t H(-ip_k,
i\bar{q_k}))}\nn\\
&&\stackrel{\lim_{N\ra\infty}}{\longrightarrow}\int_{Path}\prod_t\frac{d^nq
d^np}{(2\pi)^n}\exp{i\int^{t'}_tdt(p\cdot\dot{q}-H(-ip,iq))}
\eel{e:7}
The obvious problem with this formula is that the Hamiltonian $H(-ip, iq)$ is
{\em not} real
in general although the Hamiltonian operator $H(P,Q)$ is hermitian. Notice that
the formulas are
symmetric in the sense that \r{6} may also be written as
\be
&& \vb\phi',t'|\phi,t\hb=
\int d^nq'd^nq\phi'^*(iq')\phi(-iq)\vb -iq',t'|-iq, t\hb
\eel{e:8}
where the propagator $\vb -iq',t'|-iq, t\hb$ is given by \r{7} with $H(-ip,iq)$
replaced by
$H(ip,-iq)$ which is its complex conjugate when $H(P,Q)$ is hermitian. Hence,
$\vb\phi',t'|\phi,t\hb$ will grow
exponentially both when $t\ra\infty$ and $t\ra-\infty$ if $H(ip,-iq)$ contains
imaginary terms. To
avoid this property one has to require that $H(ip,-iq)$ is real \cite{Gross}.

Consider now hermitian fermionic canonical conjugate operators $\hpet_i$ and
$\heta_i$,
$i=1,\ldots,n$, satisfying $[\hpet_i,\heta_j|_+=\del_{ij}$. They span a finite
dimensional indefinite
metric state space. If one introduces odd Grassmann numbers one may derive
pseudoclassical path
integrals for theories involving $\hpet_i$ and $\heta_i$ and the Hamiltonian
$H(\hpet,\heta)$. In fact
there are two options: One may either make use of eigenstates with eigenvalues
which are real odd
Grassmann numbers, $\heta_i|\eta\hb=\eta_i|\eta\hb$, or which are imaginary odd
Grassmann numbers,
$\heta_i|i\eta\hb=i\eta_i|i\eta\hb$ ($\eta_i$ is real and odd). In the first
case one finds the path
integral (the conventions of \cite{Fermi} is used and appropriate orderings for
$n$ odd is ignored)
\be
&&\vb \eta',t'|\eta, t\hb=\int\prod_{m=1}^{N-1} d^n\eta_m\prod_{k=0}^{N-1}\vb
\eta_{k+1}, t_{k+1}|\eta_k, t_k\hb=\nn\\
&&=\int\prod_{m=1}^{N-1}
d^n\eta_m\prod_{k=0}^{N-1}d^n\pet_k\exp{(-\pet_k\cdot\Delta
\eta_k-i\Delta t H(\pet_k,
\bar{\eta}_k))}\nn\\
&&\stackrel{\lim_{N\ra\infty}}{\longrightarrow}\int_{Path}\prod_td^n\eta
d^n\pet\exp{i\int^{t'}_tdt(i\pet\cdot\dot{\eta}-H(\pet,\eta))}
\eel{e:81}
where $\eta_0=\eta, \;\;\eta_N,\eta_{\infty}=\eta', \;\;t_0=t, \;\;
t_N,t_{\infty}=t', \;\;\Delta
\eta_k=\eta_{k+1}-\eta_k$ and $\bar{\eta_k}=\halv(\eta_{k+1}+\eta_k)$.
$H(\pet, \eta)$ is  the
fermionic Weyl transform of the Hamiltonian operator $H(\hpet, \heta)$ defined
by
\be
&&H(\pet, \eta)=\int d^n\la d^n\xi
\tilde{H}(\la,\xi)e^{-\pet\cdot\la-\eta\cdot\xi}
\nn\\
&&H(\hpet, \heta)=\int d^n\la d^n\xi
\tilde{H}(\la,\xi)e^{-\hpet\cdot\la-\heta\cdot\xi}
\eel{e:82}
which implies
\be
&&\vb \eta''|H(\hpet,\heta)|\eta'\hb=\int d^n\pet e^{-\pet\cdot
(\eta''-\eta')}H(\pet,\bar{\eta})
\eel{e:83}
where $\bar{\eta}\equiv\halv(\eta''+\eta')$. This relation is used in \r{81}.
(The last line in
\r{81} is somewhat misleading for odd $n$ since the finite slice expression
contains one more
$\pet$-integral than $\eta$-integral which makes $\vb \eta',t'|\eta, t\hb$ odd
for odd $n$.)

Using the same conventions as above the imaginary odd Grassmann eigenstates
satisfy the properties
\be
&&\heta|i\eta\hb=i\eta|i\eta\hb,\;\;\;\vb
-i\eta|\equiv(|i\eta\hb)^{\dag},\;\;\;\vb
i\eta|i\eta'\hb=i^n\del^n(\eta-\eta')\nn\\ &&\int|i\eta\hb(-i)^n d^n\eta\vb
i\eta|=\int|-i\eta\hb i^n d^n\eta\vb -i\eta|=\bett \eel{e:84}
Instead of \r{83} we get now
\be
&&\vb i\eta''|H(\hpet,\heta)|i\eta'\hb=i^n\int d^n\pet e^{-\pet\cdot
(\eta''-\eta')}H(i\pet,-i\bar{\eta}) \eel{e:85}
which leads to the path integral
\be
&&\vb i\eta',t'|i\eta,
t\hb=\int\prod_{m=1}^{N-1}(-i)^nd^n\eta_m\prod_{k=0}^{N-1}\vb
i\eta_{k+1}, t_{k+1}|i\eta_k, t_k\hb=\nn\\
&&=\int\prod_{m=1}^{N-1}
(-i)^nd^n\eta_m\prod_{k=0}^{N-1}i^nd^n\pet_k\exp{(-\pet_k\cdot\Delta
\eta_k-i\Delta t H(i\pet_k,
-i\bar{\eta}_k))}\nn\\
&&\stackrel{\lim_{N\ra\infty}}{\longrightarrow}\int_{Path}\prod_td^n\eta
d^n\pet\exp{i\int^{t'}_tdt(i\pet\cdot\dot{\eta}-H(i\pet,-i\eta))}
\eel{e:86}
(Also this expression is misleading for odd $n$.)

 In \cite{BV} the following Hamiltonian form of the path integral for finite
dimensional bosonic
gauge theories were given (we  suppress factors of $2\pi$ in the following)
($2m<n$) \be
&&Z_{\rho}=\int d^nq d^np d^mv d^m\pi d^m\eta d^m\pet d^m\bata d^m\bapet
\times\nn\\&&\times\exp{i\int dt
(p\cdot\dot{q}+\pi\cdot\dot{v}+i\pet\cdot\dot{\eta}+i\bapet\cdot\dot{\bata}-H_{\rho})}
\eel{e:9}
where $\eta^a, \bata_a$ and $v^a$ are ghosts, antighosts and Lagrange
multipliers respectively, and
where
 \be
&&H_{\rho}\equiv H_0+\{\rho, Q\}
\eel{e:10}
is a BRST invariant Hamiltonian. $Q$ is the BRST charge, $H_0$ is BRST
invariant and $\rho$ is a
real odd gauge fixing function. In \r{10} $\rho$ has typically   the form
$\pet_av^a$ or/and $\bata_a\chi^a$.  (It must have ghost number minus one.)
Usually it turns
$H_{\rho}$ into a form which allows for an integration over the momenta in
$Z_{\rho}$ such that
\be
&&Z_{\rho}=\int D^nq D^mv D^m\eta D^m\bata \exp{i\int dt L(t)}
\eel{e:11}
where $L(t)$ is a regular configuration space Lagrangian. (Contrary to \r{9}
the configuration
space expression \r{11} usually involves a nontrivial measure.)

In the corresponding operator formulation of the  BRST quantization in \r{9}
one starts with a large
state space $\Om$ spanned by the canonical operators $(P,Q)$, $(\hat{\pi},
\hat{v})$, $(\hpet, \heta)$
and $(\hat{\bapet}, \hat{\bata})$. One defines the BRST charge operator
$\hat{Q}$ in such a way
that it is nilpotent ($\hat{Q}^2=0$), and one projects out the physical state
space $\Om_{ph}$ by
$\hat{Q}|ph\hb=0$ which is a regular state space if all zero norm states of the
form $\hat{Q}|\chi\hb$ are
divided out. What is the Hamiltonian operator? From \r{10} it is natural to
choose
\be
&&\hat{H_{\rho}}=\hat{H_0}+[\hat{\rho}, \hat{Q}]_+
\eel{e:12}
which is hermitian. This choice seems to comply with the
Fradkin-Vilkovisky theorem \cite{FV,BV,MRev} which says that the path integral
expression \r{9} is
independent of $\rho$, since the second term in \r{12} only seems to produce
zero norm states
on physical states and  which therefore may be divided out. However, this is in
general not true since
\be
&&[\hat{\rho}, \hat{Q}]_+|ph\hb=\hat{Q}\hat{\rho}|ph\hb
\eel{e:13}
is only a zero norm state {\em if} \ it also belongs to an inner product space.
 This implies that the above formal
arguments have to be replaced by more precise ones.

The operator formulation of the BRST quantization on inner product spaces was
considered in
\cite{Simple,Gauge} for gauge theories with finite number of degrees of freedom
with a nilpotent BRST
charge of the BFV form \cite{BV}
\be
&&\hat{Q}=\hat{\psi_a}\heta^a-\frac{1}{2}iU_{bc}^{\;\;a}\hpet_a
\heta^b\heta^c-\frac{1}{2}iU_{ab}^{\;\;b}\heta^a + \hat{\bapet_a}\hat{\pi}^a
\eel{e:14}
where $\hat{\psi_a}$ are hermitian bosonic gauge generators
(constraints) satisfying the Lie algebra
\be
&[\hat{\psi_a}, \hat{\psi_b}]_{-}=iU_{ab}^{\;\;c}\hat{\psi_c}
\eel{e:15}
where $U_{ab}^{\;\;c}$ are the structure constants.
By means of a bigrading \cite{RM} general solutions of $\hat{Q}|ph\hb=0$ were
derived all of the
form (apart from zero norm states)
 \be
&&|ph\hb=e^{\al[\hat{\rho}, \hat{Q}]_+}|\phi\hb
\eel{e:16}
where $|\phi\hb$ is a BRST invariant state and where $\al$ is a real parameter
different from zero.
 In \cite{Simple}
\be
&&\hat{\rho}=\hpet_a\hat{v}^a
\eel{e:17}
and  $|\phi\hb$  satisfies
\be
&&\hat{\pi}_a|\phi\hb=\heta^a|\phi\hb=\hat{\bata}_a|\phi\hb=0
\eel{e:18}
which makes $|\phi\hb$  BRST invariant.
Notice that although $|\phi\hb$ does {\em not} belong to an inner product space
$|ph\hb$ will do
provided  the quantization is appropriately prescribed. Thus, although we
formally have
\be
&&|ph\hb=|\phi\hb+\hat{Q}|\cdot\hb
\eel{e:19}
we may not divide out $\hat{Q}|\cdot\hb$. In \cite{Gauge} another set of
solutions were derived. They are
 of the form \r{16} but with
\be
&&\hat{\rho}=\hat{\bata}_a\hat{\chi^a}
\eel{e:20}
 where $\hat{\chi^a}$ is a hermitian gauge fixing operator to the gauge
generators $\hat{\psi_a}$.
($[\hat{\chi^a}, \hat{\psi_b}]_-$ must be nonsingular.) Instead of \r{18}
$|\phi\hb$ satisfies here
\be
&&\hpet_a|\phi\hb=\hat{\bapet^a}|\phi\hb=(\hat{\psi_a}+iU_{ab}^{\;\;\;b})|\phi\hb=0
\eel{e:21}
which also makes $|\phi\hb$ BRST invariant.

Consider now the physical transition amplitude
\be
&&\vb ph',t'|ph,t\hb=\vb ph'|e^{-i(t'-t)\hat{H_0}}|ph\hb=\nn\\&&=\vb
\phi'|e^{\al[\hat{\rho},
\hat{Q}]}e^{-i(t'-t)\hat{H_0}}e^{\al[\hat{\rho}, \hat{Q}]}|\phi\hb=\vb
\phi'|e^{-i(t'-t)\hat{H_0}+2\al[\hat{\rho}, \hat{Q}]}|\phi\hb \eel{e:22}
where $\hat{H_0}$ is a BRST invariant Hamiltonian operator. The last equality
is valid provided $H_0$
commutes with $[\hat{\rho}, \hat{Q}]$ which we assume. ($[\hat{\rho},
\hat{H_0}]$ must be BRST
invariant.) In \cite{Simple,Gauge} it was shown that \r{22} is independent of
the value of the real
parameter $\al$ when $t'=t$ except that it must be nonzero. This should also be
the case for $t'\neq
t$. We may therefore set $2\al=\pm(t'-t)$ except when $t'=t$. Eqn \r{22} looks
then like a transition
amplitude for the physical states $|\phi\hb$ where time translation is
generated by the non-hermitian
Hamiltonian operator
\be
&&\hat{H^c_{\rho}}=\hat{H_0}\pm i[\hat{\rho}, \hat{Q}]
\eel{e:23}
The conventional identification \r{12} leads to an imaginary $\al$ (see
\cite{Teit,Torr,MO}). which is
not allowed according to \cite{Simple,Gauge}. Eqn \r{23}  seems therefore to be
in contradiction
with the formulas \r{9} and \r{10}. However, it is at this point indefinite
metric state spaces and
imaginary eigenvalues will  help us to resolve this contradiction. In
\cite{Propa} (see  also
\cite{Gen}) quantization rules are proposed which tell us which variables may
be quantized with
indefinite metric states and which may not. For bosonic gauge theories it was
argued that the
Lagrange multipliers should be quantized with opposite metric states to the
variable which the gauge
generators $\psi_a$ eliminate. These rules lead to certain consequences for the
additional term to
the Hamiltonian $H_0$. Notice that
\be &&[\hat{\rho},
\hat{Q}]=(\hat{\psi_a}+\hat{\psi_a}^{gh})\hat{v}^a-i\hpet_a\hat{\bapet^a}\nn\\
&&\hat{\psi_a}^{gh}=\halv iU_{ab}^{\;\;\;c}(\hpet_c\heta^b-\heta^b\hpet_c)
\eel{e:24}
for $\hat{\rho}=\hat{v}^a\hpet_a$, and that
\be
&&[\hat{\rho},
\hat{Q}]=\hat{\pi}_a\hat{\chi^a}+\hat{\bata}_a\heta^b[\hat{\chi^a},
\hat{\psi_b}]
\eel{e:25}
for $\hat{\rho}=\hat{\bata}_a\hat{\chi^a}$.  We have then from \r{6} and \r{7}
that $[\hat{\rho},
\hat{Q}]$ leads to an imaginary expression in the path integral if the Lagrange
multipliers
$(\hat{\pi}, \hat{v})$ are quantized with indefinite metric states and if
imaginary eigenstates are
chosen for the fermionic ghosts $(\hpet, \heta)$ or the antighosts
$(\hat{\bapet}, \hat{\bata})$
which always is possible. Under these conditions the transition amplitude
\r{22} is consistent with
the path integral \r{9} and \r{10} provided $\hat{H_0}$ leads to a real
expression. Since the general
solutions \r{16} are expected to be valid also for graded symmetries we may
directly generalize the
above quantization rules to such theories. They imply that bosonic ghosts and
antighosts are to be
quantized with opposite metric states in agreement with the proposal in
\cite{Propa} and that
imaginary eigenstate representations are to be chosen for odd Lagrangian
multipliers. A further
implication is that theories with $H_0=0$ are invariant under time reversal
since we may choose
either sign in \r{23} and in the path integrals.

Another possibility is to consider the transition amplitude \r{22} for
imaginary times $t=-i\tau$. In
this case we may immediately identify the hermitian Hamiltonian operator \r{12}
from the choice $2\al=-(\tau'-\tau)$. It may lead to a real Hamiltonian in the
path integral provided
the Lagrange multipliers are quantized with positive metric states and if
the variable which $\psi_a$
eliminates is quantized with negative metric states and $\psi_a$ remains real.
The latter  case
occurs if \eg $\psi_a$ is quadratic in the canonical conjugate variable to the
variable which it
eliminates. This possibility is \eg used when a Euclidean propagator is derived
from the spinless
particle model in \cite{Propa}. However, particles with spin lead to complex
Hamiltonians in the path
integral. When these Euclidean propagators are analytically continued to
Minkowski space the
corresponding path integrals seem more to comply with the conventional picture
\r{9} with real
Hamiltonians. The only difference is the introduction of the
$i\epsilon$-prescription which is
necessary for convergence \cite{Teit}.

To summerize: In the operator quantization the operator $[\hat{\rho}, \hat{Q}]$
appears in order to
make the inner products finite. It is not directly connected to the Hamiltonian
operator. However, in
the transition amplitudes it  generally appears as an additional non-hermitian
term to the Hamiltonian
which due to the quantization rules that have to be imposed  appears as an
additional real term to the
Hamiltonian in the path integral. This picture requires us to quantize Lagrange
multipliers with
indefinite metric states. However, sometimes they may be quantized with
positive metric states if
instead the variables which the gauge generators $\psi_a$ eliminate are
quantized with indefinite
metric states \cite{Propa}. This is the case for propagators where the
Hamiltonians  may be
interpreted as regularized real Hamiltonians.

An important issue in the path integral expressions \r{9} and \r{11} is the
choice of boundary
conditions. As is well known good physical properties are only obtained if
certain Ward identities are
satisfied and these identities follow from the imposed boundary conditions.
The operator quantization yields apart from the above refinements of the path
integrals also a
connection between boundary conditions and the choice of gauge fixing. From the
results of
\cite{Simple,Gauge} we find that the corresponding path integral to \r{22}
requires the boundary
conditions
\be
&&\pi_a=0,\;\;\;\eta^a=0,\;\;\;\bata_a=0
\eel{e:27}
at the endpoints for $\rho=\pet_av^a$, and
\be
&&\psi_a=0,\;\;\;\pet_a=0,\;\;\;\bapet^a=0
\eel{e:28}
at the endpoints for $\rho=\bata_a\chi^a$ for unimodular gauge groups.
Possible choices of
boundary conditions were discussed in \cite{MRev} and the boundary conditions
\r{27} are identical to the ones in (1.10) there and \r{28} is partly in
agreement with (1.6-7)
($\pi_a=0$ has to be replaced  by $\bapet^a=0$). Notice that both \r{27} and
\r{28} are BRST
invariant since the $|\phi\hb$-states are BRST invariant in the general
solutions \r{14}.

\end{document}